\documentclass[a4paper,conference,twocolumn]{IEEEtran-v17}      
\usepackage[a4paper,left=15mm,right=15mm,top=25mm,bottom=25mm]{geometry}

 \usepackage{xspace}
\usepackage{url}
\usepackage[cmex10]{amsmath}
\usepackage{bbm}
\usepackage{graphicx}
\usepackage{epstopdf}
\usepackage{paralist}
\usepackage[normalem]{ulem}
\usepackage{fancyref} 
\usepackage{amsmath}
\usepackage{amssymb}

\usepackage{xcolor}
\definecolor{links}{rgb}{0.7,0,0}   
\definecolor{urls}{rgb}{0,0,0.8}    
\definecolor{cites}{rgb}{0,0,0.8}   

\usepackage{dsfont}
\usepackage{footnote}
\usepackage{balance}
\usepackage{float}
\usepackage{cite}
\usepackage{wasysym}
\usepackage{amssymb}

\usepackage{vmr-symbols-vecbold}
\usepackage{standard-macros}
\usepackage{mathbbol}

\usepackage{tikz}
\usetikzlibrary{patterns}
\usetikzlibrary{shapes,arrows,positioning}
\usetikzlibrary{calc}
\usetikzlibrary{fit}
\usetikzlibrary{plotmarks}
\tikzset{every picture/.style={font issue=\scriptsize, >=stealth},font issue/.style={execute at begin picture={#1\selectfont}}}
\tikzset{three sided left/.style={
        draw=none,
        xshift=\pgflinewidth,
        append after command={
            [shorten <= -0.5\pgflinewidth]
            ([shift={(-1.5\pgflinewidth,-0.5\pgflinewidth)}]\tikzlastnode.north east) edge ([shift={( 0.5\pgflinewidth,-0.5\pgflinewidth)}]\tikzlastnode.north west) 
            ([shift={( 0.5\pgflinewidth,-0.5\pgflinewidth)}]\tikzlastnode.north west) edge ([shift={( 0.5\pgflinewidth,+0.5\pgflinewidth)}]\tikzlastnode.south west)            
            ([shift={( 0.5\pgflinewidth,+0.5\pgflinewidth)}]\tikzlastnode.south west) edge ([shift={(-1.0\pgflinewidth,+0.5\pgflinewidth)}]\tikzlastnode.south east)
        }}}
        
\tikzset{three sided right/.style={
        draw=none,
        xshift=-\pgflinewidth,
        append after command={
            [shorten <= -0.5\pgflinewidth]
            ([shift={( 1.5\pgflinewidth,-0.5\pgflinewidth)}]\tikzlastnode.north west) edge ([shift={(-0.5\pgflinewidth,-0.5\pgflinewidth)}]\tikzlastnode.north east) 
            ([shift={(-0.5\pgflinewidth,-0.5\pgflinewidth)}]\tikzlastnode.north east) edge ([shift={(-0.5\pgflinewidth,+0.5\pgflinewidth)}]\tikzlastnode.south east)            
            ([shift={(-0.5\pgflinewidth,+0.5\pgflinewidth)}]\tikzlastnode.south east) edge ([shift={( 1.0\pgflinewidth,+0.5\pgflinewidth)}]\tikzlastnode.south west)
        }}}

\usepackage{pgfplots}
\usepgfplotslibrary{fillbetween}
\pgfplotsset{
  compat=newest, 
  width=\columnwidth,    
  height=0.8\columnwidth,   
  plot coordinates/math parser=false,
  standard/.style={
    axis equal,
    axis line style=help lines,
    axis x line=center,
    axis y line=center,
    axis z line=center},
    grid style={dashed,gray},
    minor grid style={dotted,gray},
    major grid style={dotted,gray},
    ylabel absolute, ylabel style={yshift=-0.4cm},
    xlabel absolute, xlabel style={yshift=0.25cm}
}

\makeatletter
 \pgfdeclarepatternformonly[\tikz@pattern@color,\LineSpace,\LineWidth]{my horizontal lines}%
    {\pgfpointorigin}{\pgfqpoint{100pt}{1pt}}{\pgfqpoint{100pt}{\LineSpace}}%
    {
        \pgfsetcolor{\tikz@pattern@color}
        \pgfsetlinewidth{\LineWidth}
        \pgfpathmoveto{\pgfqpoint{0pt}{0.5pt}}
        \pgfpathlineto{\pgfqpoint{100pt}{0.5pt}}
        \pgfusepath{stroke}
    }
 
 \pgfdeclarepatternformonly[\tikz@pattern@color,\LineSpace,\LineWidth]{my vertical lines}%
    {\pgfpointorigin}{\pgfqpoint{1pt}{100pt}}{\pgfqpoint{\LineSpace}{100pt}}%
    {
        \pgfsetcolor{\tikz@pattern@color}
        \pgfsetlinewidth{\LineWidth}
        \pgfpathmoveto{\pgfqpoint{0.5pt}{0pt}}
        \pgfpathlineto{\pgfqpoint{0.5pt}{100pt}}
        \pgfusepath{stroke}
    } 
 
 \pgfdeclarepatternformonly[\tikz@pattern@color,\LineSpace,\LineWidth]{my grid}%
    {\pgfqpoint{-1pt}{-1pt}}{\pgfqpoint{\LineSpace}{\LineSpace}}
    {\pgfqpoint{\LineSpace}{\LineSpace}}%
    {
        \pgfsetcolor{\tikz@pattern@color}
        \pgfsetlinewidth{\LineWidth}
        \pgfpathmoveto{\pgfqpoint{0pt}{0pt}}
        \pgfpathlineto{\pgfqpoint{0pt}{\LineSpace + 0.1pt}}
        \pgfpathmoveto{\pgfqpoint{0pt}{0pt}}
        \pgfpathlineto{\pgfqpoint{\LineSpace + 0.1pt}{0pt}}
        \pgfusepath{stroke}
    }
 
 \pgfdeclarepatternformonly[\tikz@pattern@color,\LineSpace,\LineWidth]{my north east lines}
    {\pgfqpoint{-\LineWidth}{-\LineWidth}}{\pgfqpoint{\LineSpace}{\LineSpace}}
    {\pgfqpoint{\LineSpace}{\LineSpace}}%
    {
        \pgfsetcolor{\tikz@pattern@color}
        \pgfsetlinewidth{\LineWidth}
        \pgfpathmoveto{\pgfqpoint{-\LineWidth}{-\LineWidth}}
        \pgfpathlineto{\pgfqpoint{\LineSpace + 0.1pt}{\LineSpace + 0.1pt}}
        \pgfusepath{stroke}
    }
 
 \pgfdeclarepatternformonly[\tikz@pattern@color,\LineSpace,\LineWidth]{my north west lines}
    {\pgfqpoint{-\LineWidth}{-\LineWidth}}{\pgfqpoint{\LineSpace}{\LineSpace}}
    {\pgfqpoint{\LineSpace}{\LineSpace}}%
    {
        \pgfsetcolor{\tikz@pattern@color}
        \pgfsetlinewidth{\LineWidth}
        \pgfpathmoveto{\pgfqpoint{-\LineWidth}{\LineSpace}}
        \pgfpathlineto{\pgfqpoint{\LineSpace + 0.1pt}{-\LineWidth}}
        \pgfusepath{stroke}
    }
 
 \pgfdeclarepatternformonly[\tikz@pattern@color,\LineSpace,\LineWidth]{my crosshatch}%
    {\pgfqpoint{-1pt}{-1pt}}{\pgfqpoint{\LineSpace}{\LineSpace}}
    {\pgfqpoint{\LineSpace}{\LineSpace}}%
    {
        \pgfsetcolor{\tikz@pattern@color}
        \pgfsetlinewidth{\LineWidth}
        \pgfpathmoveto{\pgfqpoint{\LineSpace + 0.1pt}{0pt}}
        \pgfpathlineto{\pgfqpoint{0pt}{\LineSpace + 0.1pt}}
        \pgfpathmoveto{\pgfqpoint{0pt}{0pt}}
        \pgfpathlineto{\pgfqpoint{\LineSpace + 0.1pt}{\LineSpace + 0.1pt}}
        \pgfusepath{stroke}
    }
 
 \pgfdeclarepatternformonly[\tikz@pattern@color,\LineSpace,\PointSize]{my dots}%
    {\pgfqpoint{-\LineSpace*0.25}{-\LineSpace*0.25}}
    {\pgfqpoint{\LineSpace*0.25}{\LineSpace*0.25}}
    {\pgfqpoint{\LineSpace*0.75}{\LineSpace*0.75}}%
    {
        \pgfsetcolor{\tikz@pattern@color}
        \pgfpathcircle{\pgfqpoint{0pt}{0pt}}{\PointSize}
        \pgfusepath{fill}
    }
\makeatother
 
\newdimen\LineSpace
\newdimen\PointSize
\newdimen\LineWidth
\tikzset{
    line space/.code={\LineSpace=#1},
    line space=3pt
}
\tikzset{
    point size/.code={\PointSize=#1},
    point size=.5pt
}
\tikzset{
    pattern line width/.code={\LineWidth=#1},
    pattern line width=.4pt
}

\DeclareSymbolFontAlphabet{\amsmathbb}{AMSb}%

\newcommand{\lro}[1]{\lefto({#1}\right)}																
\newcommand{\lrbo}[1]{\lefto \lbrace {#1} \right \rbrace}															
\newcommand{\lrho}[1]{\lefto [ {#1} \right ]}																				

\newcommand{\lr}[1]{\left({#1}\right)}																
\newcommand{\lrb}[1]{\left \lbrace {#1} \right \rbrace}															
\newcommand{\lrh}[1]{\left [ {#1} \right ]}																				

\safemath{\dopplerspread}{B_D}																								
\safemath{\delayspread}{T_D}																									
\safemath{\nc}{n\sub{c}}																										
\safemath{\nd}{n\sub{d}}																										
\safemath{\ntx}{n\sub{t}} 																											
\safemath{\nrx}{n\sub{r}}																											
\safemath{\ntxt}{\tilde{n\sub{t}}}																											
\safemath{\cb}{\ensuremath{L}} 																								
\safemath{\cl}{\ensuremath{n}} 																								
\safemath{\txanto}{{\ensuremath{\tilde{m}_t}}} 																		
\safemath{\cs}{M} 																														
\safemath{\idPustm}{\ensuremath{S_{k}}}
\safemath{\error}{\ensuremath{\epsilon}} 																				
\safemath{\eexp}{\ensuremath{\mathcal{E}}} 																			
\safemath{\nsubc}{n\sub{s}}			 																						
\safemath{\nofdm}{n\sub{o}} 																									
\safemath{\bc}{\ensuremath{B_c}} 																							
\safemath{\ts}{\ensuremath{T_s}} 																							
\safemath{\nrb}{\ensuremath{n_{rb}}} 																						
\safemath{\nres}{\ell}
\newcommand{\cgauss}[2]{\mathcal{CN}\lro{\ensuremath{#1, #2}  }}   								
\safemath{\maxk}{M^*\lr{\nres, \nsubc, \nofdm, \epsilon, \rho}}
\safemath{\Rmax}{R^*}
\safemath{\Emin}{E\sub{b}^*/N_0}
\safemath{\Eminf}{\frac{E\sub{b}^*}{N_0}}
\safemath{\np}{\ensuremath{n\sub{p}}}
\safemath{\code}{\ensuremath{\mathcal{C}}}
\safemath{\err}{\ensuremath{\epsilon}}
\safemath{\rp}{\ensuremath{\rho\sub{p}}}
\safemath{\rd}{\ensuremath{\rho\sub{d}}}
\safemath{\cohtime}{\ensuremath{T\sub{c}}}
\safemath{\cohbw}{\ensuremath{B\sub{c}}}
\safemath{\nmax}{\ensuremath{\ell_{\text{max}}}}

\safemath{\xdd}{\ensuremath{\vecx^{(\text{d})}}}
\safemath{\xp}{\ensuremath{\vecx^{(\text{p})}}}
\safemath{\xd}{\ensuremath{\randvecx^{(\text{d})}}}
\safemath{\yp}{\ensuremath{\randvecy^{(\text{p})}}}
\safemath{\yd}{\ensuremath{\randvecy^{(\text{d})}}}

\safemath{\matpilotbar}{\ensuremath{\overline{\randvecx}^{(\text{p})}}}
\safemath{\matdatabar}{\ensuremath{\overline{\randvecx}^{(\text{d})}}}

\newcommand{\prob}[1]{\ensuremath{\mathbb{P}\lrho{#1}}}

\safemath{\mI}{\ensuremath{i\lro{\randvecy ; \randvecx}}} 				

\safemath{\randveca}{\bm{A}}
\safemath{\randvecb}{\bm{B}}
\safemath{\randvecc}{\bm{C}}
\safemath{\randvecd}{\bm{D}}
\safemath{\randvece}{\bm{E}}
\safemath{\randvecf}{\bm{F}}
\safemath{\randvecg}{\bm{G}}
\safemath{\randvech}{\bm{H}}
\safemath{\randveci}{\bm{I}}
\safemath{\randvecj}{\bm{J}}
\safemath{\randveck}{\bm{K}}
\safemath{\randvecl}{\bm{L}}
\safemath{\randvecm}{\bm{M}}
\safemath{\randvecn}{\bm{N}}
\safemath{\randveco}{\bm{O}}
\safemath{\randvecp}{\bm{P}}
\safemath{\randvecq}{\bm{Q}}
\safemath{\randvecr}{\bm{R}}
\safemath{\randvecs}{\bm{S}}
\safemath{\randvect}{\bm{T}}
\safemath{\randvecu}{\bm{U}}
\safemath{\randvecv}{\bm{V}}
\safemath{\randvecw}{\bm{W}}
\safemath{\randvecx}{\bm{X}}
\safemath{\randvecy}{\bm{Y}}
\safemath{\randvecz}{\bm{Z}}
\safemath{\randvecphi}{\bm{\Phi}}

\safemath{\randmatA}{\amsmathbb{A}}
\safemath{\randmatB}{\amsmathbb{B}}
\safemath{\randmatC}{\amsmathbb{C}}
\safemath{\randmatD}{\amsmathbb{D}}
\safemath{\randmatE}{\amsmathbb{E}}
\safemath{\randmatF}{\amsmathbb{F}}
\safemath{\randmatG}{\amsmathbb{G}}
\safemath{\randmatH}{\amsmathbb{H}}
\safemath{\randmatI}{\amsmathbb{I}}
\safemath{\randmatJ}{\amsmathbb{J}}
\safemath{\randmatK}{\amsmathbb{K}}
\safemath{\randmatL}{\amsmathbb{L}}
\safemath{\randmatM}{\amsmathbb{M}}
\safemath{\randmatN}{\amsmathbb{N}}
\safemath{\randmatO}{\amsmathbb{O}}
\safemath{\randmatP}{\amsmathbb{P}}
\safemath{\randmatQ}{\amsmathbb{Q}}
\safemath{\randmatR}{\amsmathbb{R}}
\safemath{\randmatS}{\amsmathbb{S}}
\safemath{\randmatT}{\amsmathbb{T}}
\safemath{\randmatU}{\amsmathbb{U}}
\safemath{\randmatV}{\amsmathbb{V}}
\safemath{\randmatW}{\amsmathbb{W}}
\safemath{\randmatX}{\amsmathbb{X}}
\safemath{\randmatY}{\amsmathbb{Y}}
\safemath{\randmatZ}{\amsmathbb{Z}}
\safemath{\randmatSigma}{\mathbb{\Sigma}}
\safemath{\randmatPhi}{\mathbb{\Phi}}
\safemath{\randmatLambda}{\mathbb{\Lambda}}

\safemath{\matSigma}{\bm{\Sigma}}
\safemath{\matPhi}{\bm{\Phi}}
\safemath{\matLambda}{\bm{\Lambda}}

\usepackage{glossaries}
\makeglossaries

\newglossaryentry{3gpp}
{
  name={3GPP},
  description={3rd generation partnership project},
  first={\glsentrydesc{3gpp} (\glsentrytext{3gpp})}
} 

\newglossaryentry{dl}
{
  name={DL},
  description={downlink},
  first={\glsentrydesc{dl} (\glsentrytext{dl})}
} 

\newglossaryentry{LED}
{
  name={LED},
  description={light emitting diode},
  first={\glsentrydesc{LED} (\glsentrytext{LED})},
  plural={LEDs},
  descriptionplural={light emitting diodes},
  firstplural={\glsentrydescplural{LED} (\glsentryplural{LED})}
} 

\newglossaryentry{lte}
{
  name={LTE},
  description={long-term evolution},
  first={\glsentrydesc{lte} (\glsentrytext{lte})}
} 
\newglossaryentry{ofdm}
{
  name={OFDM},
  description={orthogonal frequency-division multiplexing},
  first={\glsentrydesc{ofdm} (\glsentrytext{ofdm})}
} 
\newglossaryentry{ue}
{
  name={UE},
  description={user equipment},
  first={\glsentrydesc{ue} (\glsentrytext{ue})},
  plural={UE's},
  descriptionplural={user equipments},
  firstplural={\glsentrydescplural{ue} (\glsentryplural{ue})}
} 
\newglossaryentry{ul}
{
  name={UL},
  description={uplink},
  first={\glsentrydesc{ul} (\glsentrytext{ul})}
} 
\newglossaryentry{rb}
{
  name={RB},
  description={resource block},
  first={\glsentrydesc{rb} (\glsentrytext{rb})},
  plural={RBs},
  descriptionplural={resource blocks},
  firstplural={\glsentrydescplural{rb} (\glsentryplural{rb})}
} 
\newglossaryentry{cu}
{
  name={CU},
  description={channel use},
  first={\glsentrydesc{cu} (\glsentrytext{cu})},
  plural={CUs},
  descriptionplural={channel uses},
  firstplural={\glsentrydescplural{cu} (\glsentryplural{cu})}
} 
\newglossaryentry{tti}
{
  name={TTI},
  description={transmission time interval},
  first={\glsentrydesc{tti} (\glsentrytext{tti})},
  plural={TTI's},
  descriptionplural={transmission time intervals},
  firstplural={\glsentrydescplural{tti} (\glsentryplural{tti})}
} 
\newglossaryentry{iid}
{
  name={i.i.d.},
  description={identical and independently distributed},
  first={\glsentrydesc{iid} (\glsentrytext{iid})}
} 
\newglossaryentry{psd}
{
  name={PSD},
  description={power spectral density},
  first={\glsentrydesc{psd} (\glsentrytext{psd})}
} 

\newglossaryentry{mimo}
{
  name={MIMO},
  description={multiple-input multiple-output},
  first={\glsentrydesc{mimo} (\glsentrytext{mimo})}
} 

\newglossaryentry{bler}
{
  name={BLER},
  description={block error probability},
  first={\glsentrydesc{bler} (\glsentrytext{bler})}
} 
\newglossaryentry{mtc}
{
  name={MTC},
  description={machine-type communication},
  first={\glsentrydesc{mtc} (\glsentrytext{mtc})}
}
\newglossaryentry{csi}
{
  name={CSI},
  description={channel state information},
  first={\glsentrydesc{csi} (\glsentrytext{csi})}
}
\newglossaryentry{dvbt}
{
  name={DVB-T},
  description={terrestrial digital video broadcasting},
  first={\glsentrydesc{dvbt} (\glsentrytext{dvbt})}
}
\newglossaryentry{pat}
{
  name={PAT},
  description={pilot-assisted transmission},
  first={\glsentrydesc{pat} (\glsentrytext{pat})}
}
\newglossaryentry{ml}
{
  name={ML},
  description={maximum likelihood},
  first={\glsentrydesc{ml} (\glsentrytext{ml})}
}
\newglossaryentry{dt}
{
  name={DT},
  description={dependence-testing},
  first={\glsentrydesc{dt} (\glsentrytext{dt})}
}
\newglossaryentry{ustm}
{
  name={USTM},
  description={unitary space-time modulation},
  first={\glsentrydesc{ustm} (\glsentrytext{ustm})}
}
\newglossaryentry{siso}
{
  name={SISO},
  description={single-input single-output},
  first={\glsentrydesc{siso} (\glsentrytext{siso})}
}
\newglossaryentry{snn}
{
  name={SNN},
  description={scaled nearest-neighbor},
  first={\glsentrydesc{snn} (\glsentrytext{snn})}
}
\newglossaryentry{rcus}
{
  name={RCUs},
  description={random-coding union bound with parameter $s$},
  first={\glsentrydesc{rcus} (\glsentrytext{rcus})}
}
\newglossaryentry{osd}
{
  name={OSD},
  description={ordered statistics decoding},
  first={\glsentrydesc{osd} (\glsentrytext{osd})}
}
\newglossaryentry{llr}
{
  name={LLR},
  description={log-likelihood ratio},
  first={\glsentrydesc{llr} (\glsentrytext{llr})}
   plural={LLR's},
  descriptionplural={log-likelihood ratios},
  firstplural={\glsentrydescplural{llr} (\glsentryplural{llr})}
}
\newglossaryentry{qpsk}
{
  name={QPSK},
  description={quaternary phase-shift keying},
  first={\glsentrydesc{qpsk} (\glsentrytext{qpsk})}
}
\newglossaryentry{nlos}
{
  name={NLOS},
  description={non-line-of-sight},
  first={\glsentrydesc{nlos} (\glsentrytext{nlos})}
}
\newglossaryentry{los}
{
  name={LOS},
  description={line-of-sight},
  first={\glsentrydesc{los} (\glsentrytext{los})}
}
\newglossaryentry{cgf}
{
  name={CGF},
  description={cumulant-generating function},
  first={\glsentrydesc{cgf} (\glsentrytext{cgf})}
   plural={CGF's},
  descriptionplural={cumulant-generating functions},
  firstplural={\glsentrydescplural{cgf} (\glsentryplural{cgf})}
}
\newglossaryentry{pdf}
{
  name={PDF},
  description={probability density function},
  first={\glsentrydesc{pdf} (\glsentrytext{pdf})}
   plural={PDF's},
  descriptionplural={probability density functions},
  firstplural={\glsentrydescplural{pdf} (\glsentryplural{pdf})}
}
\newglossaryentry{gmi}
{
  name={GMI},
  description={generalized mutual information},
  first={\glsentrydesc{gmi} (\glsentrytext{gmi})}
}

\newglossaryentry{harqir}
{
  name={HARQ},
  description={hybrid automatic repeat request},
  first={\glsentrydesc{harqir} (\glsentrytext{harqir})}
}

\newglossaryentry{fbl}
{
  name={FBL-NF},
  description={fixed blocklength and no feedback},
  first={\glsentrydesc{fbl} (\glsentrytext{fbl})}
}
\newglossaryentry{ssnf}
{
  name={SS-NF},
  description={single-shot no-feedback},
  first={\glsentrydesc{ssnf} (\glsentrytext{ssnf})}
}

\newglossaryentry{urllc}
{
  name={URLLC},
  description={ultra-reliable low-latency communications},
  first={\glsentrydesc{urllc} (\glsentrytext{urllc})}
}

\newglossaryentry{vlsf}
{
  name={VLSF},
  description={variable-length stop-feedback},
  first={\glsentrydesc{vlsf} (\glsentrytext{vlsf})}
}

\newglossaryentry{cdf}
{
  name={CDF},
  description={cumulative distribution function},
  first={\glsentrydesc{cdf} (\glsentrytext{cdf})}
}

\newglossaryentry{ccdf}
{
  name={CCDF},
  description={complementary cumulative distribution function},
  first={\glsentrydesc{ccdf} (\glsentrytext{ccdf})}
}
\usepackage{pgfplots}
\usepgfplotslibrary{groupplots}
\usetikzlibrary{pgfplots.groupplots}
\usepgfplotslibrary{external}

\newcommand\linew{1pt} 

\makeatletter
\def\@IEEEinterspaceratioM{0.265}
\def\@IEEEinterspaceMINratioM{0.1651}
\def\@IEEEinterspaceMAXratioM{0.38}

\def\@IEEEinterspaceratioB{0.31}
\def\@IEEEinterspaceMINratioB{0.19}
\def\@IEEEinterspaceMAXratioB{0.38}
\@IEEEtunefonts
\makeatother
\let\abs\undefined
\newcommand{\abs}[1]{\lvert#1\rvert}		

\definecolor{red}{RGB}{220, 10, 10}
\definecolor{blue}{RGB}{10, 40, 200}
\definecolor{green}{RGB}{10, 200, 10}
\definecolor{orange}{RGB}{255, 165, 0}
\definecolor{pink}{RGB}{255, 51, 204}

\pgfplotscreateplotcyclelist{colorbrewer}{
{red},
{blue},
{green},
{orange},
{pink}
}
\pgfplotscreateplotcyclelist{colorfour}{
{red},
{blue},
{green},
{orange}
}

\makeatletter
\let\MYcaption\@makecaption
\makeatother

\usepackage[font=scriptsize]{subcaption}
\makeatletter
\let\@makecaption\MYcaption
\makeatother

\definecolor{rahulcommentcolor}{HTML}{0303BF}

\let\marginnote\undefined

\newcommand{\marginnote}[1]{\marginpar{\framebox{\framebox{#1}}}}

\newcommand\blfootnote[1]{%
  \begingroup
  \renewcommand\thefootnote{}\footnote{#1}%
  \addtocounter{footnote}{-1}%
  \endgroup
}

\usepackage[scaled=0.9]{helvet}
\usepackage{tabularx,booktabs}
\newcolumntype{L}[1]{>{\raggedright\let\newline\\\arraybackslash\hspace{0pt}}m{#1}}
\newcolumntype{C}[1]{>{\centering\let\newline\\\arraybackslash\hspace{0pt}}m{#1}}
\newcolumntype{R}[1]{>{\raggedleft\let\newline\\\arraybackslash\hspace{0pt}}m{#1}}
\usepackage{boldline}

\begin{document}
\title{Low-Latency Short-Packet Transmissions: \\ Fixed Length or HARQ?}
  \author{\IEEEauthorblockN{Johan \"Ostman, Rahul Devassy, Guido Carlo Ferrante, Giuseppe Durisi\\
  Chalmers University of Technology,  Gothenburg, Sweden}}

\maketitle

\begin{abstract}
\boldmath
We study short-packet communications, subject to latency and reliability constraints, under the premises of limited frequency diversity and no time diversity.
The question addressed is whether, and when, hybrid automatic repeat request (HARQ) outperforms fixed-blocklength schemes with no feedback (FBL-NF)  in such a setting.
We derive an achievability bound for HARQ, under the assumption of a limited number of transmissions. 
The bound relies on pilot-assisted transmission to estimate the fading channel and scaled nearest-neighbor decoding at the receiver.
We compare our achievability bound for HARQ to state-of-the-art achievability bounds for FBL-NF communications and show that for a given latency, reliability, number of information bits, and number of diversity branches, HARQ may significantly outperform FBL-NF.
For example, for an average latency of $1$ ms, a target error probability of $10^{-3}$, $30$ information bits, and  $3$ diversity branches, the gain in energy per bit is about $4$ dB.\blfootnote{This work was supported by the Swedish Research Council under grants 2014-6066 and 2016-03293.}
\end{abstract}

\section{Introduction}
\label{sec:introduction}

Fifth generation (5G) wireless systems are envisioned to support communications with stringent requirements on reliability and latency---the so called \gls{urllc}~\cite{itu-10-17}.
The objective is to enable use cases like smart grids and wireless industrial control. %
While previous-generation wireless systems focused mostly on the support of large data rates, \gls{urllc} systems target transmission of small data payloads, carried by short coded packets to meet the latency requirements.

Performance analyses of wireless systems are often carried out using asymptotic information-theoretic metrics like ergodic capacity and outage capacity, under the implicit assumption that a message is encoded into an arbitrarily large number of coded symbols. %
Although a good approximation for systems where latency concerns are secondary, such performance analyses are not suitable for \gls{urllc}~\cite{durisi16-09a}.
Instead, to study the performance of \gls{urllc} links, one must resort to tools that allow for the analysis of communication systems in the nonasymptotic regime where a message is encoded onto a finite, often small, number of symbols. %

For communications with \gls{fbl}, the nonasymptotic information-theoretic tools developed in \cite{polyanskiy10-05a} have recently enabled the characterization of the maximum coding rate achievable for a given blocklength and error probability in many scenarios of practical relevance for 5G and beyond. 
For example, it has allowed for the study of the rate achievable with short packets on general quasi-static fading channels\cite{yang14-07c} and \gls{mimo} Rayleigh block-fading channels~\cite{durisi16-02a}.
Furthermore, the practically relevant case of \gls{pat} and \gls{snn} decoding has been analyzed for \gls{siso} Rician block-fading channels~\cite{johan:unpublished} and for \gls{mimo} Rayleigh block-fading channels~\cite{ferrante18}.

Finite-blocklength information theory prescribes that for several channels, including the AWGN channel, short-packet transmission incurs a rate loss from capacity roughly proportional to the inverse of the square root of the blocklength~\cite{polyanskiy10-05a}.
However, when one is allowed to use \gls{vlsf} codes, the picture is different.
Specifically, capacity is approached much faster in the blocklength~\cite{polyanskiy2011feedback}.
\gls{vlsf} codes are an instance of variable-length feedback codes where the encoder keeps on transmitting coded symbols until it receives a single-bit feedback sent by the decoder to inform the encoder that decoding is complete.
Incremental redundancy \gls{harqir} is an instance of \gls{vlsf} codes.
In~\cite{polyanskiy2011feedback}, decoding is attempted upon reception of every new symbol.
This setup was later extended to blocks of symbols in~\cite{williamson15-07a}, a scenario that is more relevant for \gls{harqir}.
Under the assumption that the received symbols are discarded and transmission starts over if decoding is not successful after a finite number of blocks, the rate penalty resulting from decoding after each block of symbols rather than after each symbol was characterized in~\cite{williamson15-07a} for the AWGN channel.
However, such strategy still outperforms \gls{fbl} codes for moderate latencies.

In~\cite{Avranas18}, \gls{fbl} and \gls{harqir} are compared for the AWGN channel by taking into account the delay caused by the transmission of a positive/negative acknowledgement (ACK/NACK).
The analysis, however, does not account for the event of an undetected error, i.e., the event that an ACK is fed back although the decision of the decoder is wrong.
Furthermore, the normal approximation~\cite{polyanskiy10-05a} is used in the analysis.
However, this approximation, which is based on the Berry-Esseen theorem, is accurate only when the blocklength and the error probability are not too small and may not be suitable for \gls{urllc}.

\paragraph*{Contributions}
We consider an \gls{harqir} scheme employing \gls{pat} and \gls{snn} decoding at the receiver for transmission of short data packets over a \gls{siso} block-fading channel.
Leveraging~\cite[Thm. 3]{polyanskiy2011feedback}, which provides an extension of the dependence-testing bound~\cite[Thm. 17]{polyanskiy10-05a} for \gls{fbl} to \gls{vlsf} codes, we derive an achievability bound on the minimum energy per bit required to transmit a small information payload under a given latency and reliability target.
Differently from \cite[Thm. 3]{polyanskiy2011feedback}, where decoding stops when the accumulated information density corresponding to one of the codewords exceeds a threshold, in our setup decoding stops when the accumulated \emph{generalized} information density exceeds a threshold or the number of transmissions reaches a maximum predetermined value.

Our bound depends on the \gls{cdf} of the decoding time, i.e., the time at which the decoder stops and makes a decision.
This quantity has recently been shown to be central in joint coding-queuing analyses of the delay-violation probability in \gls{urllc} systems~\cite{rahul:unpublished}.
Our analysis provides  means to assess whether \gls{harqir} results in lower minimum energy per bit and higher maximum coding rate than \gls{fbl}.
In particular, our numerical results show that \gls{harqir} may significantly outperform \gls{fbl} for low-latency targets.

\paragraph*{Notation}
We denote random vectors and scalars by uppercase bold and standard letters, such as $\randvecx$ and $X$, respectively, and their realizations by lower-case letters.
The identity matrix of size $a\times a$ is written as $\rmatI_{a}$.
The distribution of a circularly-symmetric complex Gaussian random variable with variance $\sigma^2$ is denoted by $\cgauss{0}{\sigma^2}$.
The superscript~$\herm{\lro{\cdot }}$ denotes Hermitian transposition.
We write $\log(\cdot)$ and $\log_2(\cdot)$ to denote the natural logarithm and the logarithm to the base $2$, respectively.
Finally, $\lrh{a}^+$ stands for $\max\lrbo{0, a}$, $\vecnorm{\cdot}$ denotes the Euclidean norm, and $\Ex{}{\cdot}$ is the expectation operator.

\section{System Model} 
\label{sec:system_model}

\subsection{Setup}
We consider a memoryless \gls{siso} Rayleigh block-fading channel in which fading is assumed to stay constant within a coherence block and to change independently across coherence blocks.
The channel coherence time and coherence bandwidth are denoted by $\cohtime$ and $\cohbw$, respectively. %
For a system bandwidth~$B$, the available number of \emph{diversity branches} is $L\sub{c} = \floor{B/\cohbw}$. %
The duration of a transmitted codeword is assumed to be much smaller than the coherence time.
Hence, no time diversity is available.
This is in line with commonly used fourth generation (4G) and 5G channel models, as discussed in Section~\ref{sec:numerical_results}.
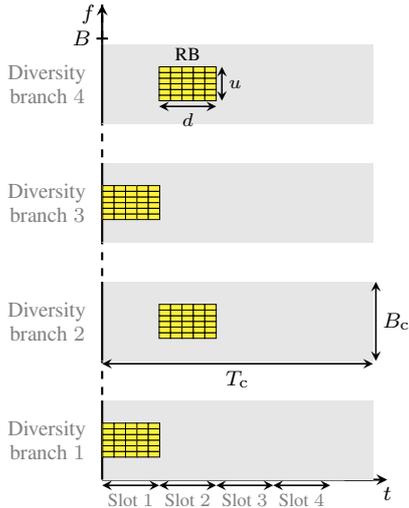
\begin{figure}[t]
\centering
\begin{tikzpicture}[scale=0.75]
    

    \draw[thick]  (0,0)--(4.75,0);
    \draw[->,>=stealth,thick] (4.25,0) -- (5,0) node[anchor=north] {\footnotesize $t$};

    \foreach \y [count=\yi] in {0,2.1 ,  4.2,6.3}  {
    	\fill[color=gray!20!white] (0,\y) rectangle (4.75,\y+1.4);
    	\draw[thick] (0,\y) -- (0,\y+1.4);
    	\draw[thick,dashed] (0,\y+1.5) -- (0,\y+2.1);
    	\node[anchor=east, gray] at (0,\y + 0.7) {\footnotesize \parbox{10ex}{\centering Diversity \\ branch $\yi$}};
    }
  
    \draw[->,>=stealth,thick]  (0,7.7)--(0,8.4) node[near end,anchor=east]{\footnotesize $f$};
    \draw[thick] (-0.1,7.8) -- (0.1,7.8) node [anchor=east]{\footnotesize $B\hspace{1ex}$};

    \foreach \x in {0,1,2,3,4} {
    	\foreach \y in {0,1,2,3,4,5} {
    	    \fill[color=yellow!85, draw=black,very thin] (0.01+0.2*\x, 0.4+\y*0.1) rectangle (0.01+0.2*\x+0.2,0.4 +\y*0.1+0.1);
	}
    }    
    
     
    \foreach \x in {0,1,2,3,4} {
    	\foreach \y in {0,1,2,3,4,5} {
    	    \fill[color=yellow!85, draw=black,very thin] (1+0.2*\x, 2.5+\y*0.1) rectangle (1+0.2*\x+0.2,2.5 +\y*0.1+0.1);
	}
    }     
 

        \foreach \x in {0,1,2,3,4} {
    	\foreach \y in {0,1,2,3,4,5} {
    	    \fill[color=yellow!85, draw=black,very thin] (0.01+0.2*\x, 4.6+\y*0.1) rectangle (0.01+0.2*\x+0.2,4.6 +\y*0.1+0.1);
	}
    }   
     
             \foreach \x in {0,1,2,3,4} {
    	\foreach \y in {0,1,2,3,4,5} {
    	    \fill[color=yellow!85, draw=black,very thin] (1+0.2*\x,6.7+\y*0.1) rectangle (1+0.2*\x+0.2,6.7 +\y*0.1+0.1);
	}
    }   
    

   \draw[semithick,<->,>=stealth] (0,-0.1) -- +(1,0) node[midway, anchor=north, inner sep=0,yshift=-0.1cm, gray] {\scriptsize Slot $1$};
   \draw[semithick,<->,>=stealth] (1,-0.1) -- +(1,0) node[midway, anchor=north, inner sep=0,yshift=-0.1cm, gray] {\scriptsize Slot $2$};
   \draw[semithick,<->,>=stealth] (2,-0.1) -- +(1,0) node[midway, anchor=north, inner sep=0,yshift=-0.1cm, gray] {\scriptsize Slot $3$};
   \draw[semithick,<->,>=stealth] (3,-0.1) -- +(1,0) node[midway, anchor=north, inner sep=0,yshift=-0.1cm, gray] {\scriptsize Slot $4$};

    \draw[semithick,<->,>=stealth] (4.8,2.1)-- +(0,1.4) node[midway, right] {\footnotesize $B\sub{c}$};
    \draw[semithick,<->,>=stealth] (0,2.05) -- +(4.75,0) node[midway, anchor=north] {\footnotesize $T\sub{c}$};

%
    
    \draw[semithick,<->,>=stealth] (1,6.6) -- +(1,0) node[midway, anchor=north] {\scriptsize $d$};
    \draw[semithick,<->,>=stealth] (2.1,6.7) -- +(0,0.6) node[midway, anchor=west] {\scriptsize $u$};    
    \node[anchor=south] at (1.5,7.3) {\scriptsize  RB};

    
\end{tikzpicture}
\caption{Example of the \gls{fbl} scheme with parameters $u=6$, $d=5$, $L\sub{c}=4$, $L=2$, and $v=2$.}
 \label{fig:FBL}
\end{figure}
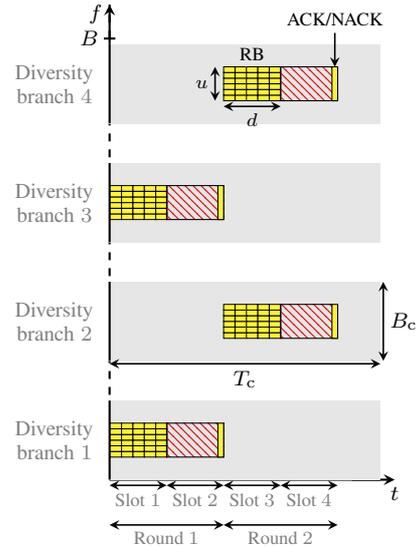
\begin{figure}[t]
\centering
\begin{tikzpicture}[scale=0.75]
    

    \draw[thick]  (0,0)--(4.75,0);
    \draw[->,>=stealth,thick] (4.25,0) -- (5,0) node[anchor=north] {\footnotesize $t$};

    \foreach \y [count=\yi] in {0,2.1 ,  4.2,6.3}  {
    	\fill[color=gray!20!white] (0,\y) rectangle (4.75,\y+1.4);
    	\draw[thick] (0,\y) -- (0,\y+1.4);
    	\draw[thick,dashed] (0,\y+1.5) -- (0,\y+2.1);
    	\node[anchor=east, gray] at (0,\y + 0.7) {\footnotesize \parbox{10ex}{\centering Diversity \\ branch $\yi$}};
    }
  
    \draw[->,>=stealth,thick]  (0,7.7)--(0,8.4) node[near end,anchor=east]{\footnotesize $f$};
    \draw[thick] (-0.1,7.8) -- (0.1,7.8) node [anchor=east]{\footnotesize $B\hspace{1ex}$};

    \foreach \x in {0,1,2,3,4} {
    	\foreach \y in {0,1,2,3,4,5} {
    	    \fill[yellow!85, draw=black,very thin] (0.01+0.2*\x, 0.4+\y*0.1) rectangle (0.01+0.2*\x+0.2,0.4 +\y*0.1+0.1);
	}
    }    
    \draw[pattern=north west lines, pattern color=red] (1,0.4) rectangle +(1,0.6);
	\draw[draw = black, fill = yellow!85] (1.9,0.4) rectangle +(0.1,0.6);
     
    \foreach \x in {0,1,2,3,4} {
    	\foreach \y in {0,1,2,3,4,5} {
    	    \fill[color=yellow!85, draw=black,very thin] (2+0.2*\x, 2.5+\y*0.1) rectangle (2+0.2*\x+0.2,2.5 +\y*0.1+0.1);
	}
    }     
     \draw[pattern=north west lines, pattern color=red] (3,2.5) rectangle +(1,0.6);
	\draw[draw = black, fill = yellow!85]  (3.9,2.5) rectangle +(0.1,0.6);
    
        \foreach \x in {0,1,2,3,4} {
    	\foreach \y in {0,1,2,3,4,5} {
    	    \fill[color=yellow!85, draw=black,very thin] (0.01+0.2*\x, 4.6+\y*0.1) rectangle (0.01+0.2*\x+0.2,4.6 +\y*0.1+0.1);
	}
    }   
     \draw[pattern=north west lines, pattern color=red] (1,4.6) rectangle +(1,0.6);
	\draw[draw = black, fill = yellow!85]  (1.9,4.6) rectangle +(0.1,0.6);
     
             \foreach \x in {0,1,2,3,4} {
    	\foreach \y in {0,1,2,3,4,5} {
    	    \fill[color=yellow!85, draw=black,very thin] (2+0.2*\x,6.7+\y*0.1) rectangle (2+0.2*\x+0.2,6.7 +\y*0.1+0.1);
	}
    }   
    \draw[pattern=north west lines, pattern color=red] (3,6.7) rectangle +(1,0.6);
    \draw[draw = black, fill = yellow!85]  (3.9,6.7) rectangle +(0.1,0.6);
    
    \draw[->,semithick]  (3.95,7.9)--(3.95,7.3) node[inner sep = 0pt] at (3.95,8.15) {\scriptsize ACK/NACK};
    
   \draw[semithick,<->,>=stealth] (0,-0.1) -- +(1,0) node[midway, anchor=north, inner sep=0,yshift=-0.1cm, gray] {\scriptsize Slot $1$};
   \draw[semithick,<->,>=stealth] (1,-0.1) -- +(1,0) node[midway, anchor=north, inner sep=0,yshift=-0.1cm, gray] {\scriptsize Slot $2$};
   \draw[semithick,<->,>=stealth] (2,-0.1) -- +(1,0) node[midway, anchor=north, inner sep=0,yshift=-0.1cm, gray] {\scriptsize Slot $3$};
   \draw[semithick,<->,>=stealth] (3,-0.1) -- +(1,0) node[midway, anchor=north, inner sep=0,yshift=-0.1cm, gray] {\scriptsize Slot $4$};
   
      \draw[semithick,<->,>=stealth] (0,-0.8) -- +(2,0) node[midway, anchor=north, inner sep=0,yshift=-0.075cm, gray] {\scriptsize Round $1$};
   \draw[semithick,<->,>=stealth] (2,-0.8) -- +(2,0) node[midway, anchor=north, inner sep=0,yshift=-0.075cm, gray] {\scriptsize Round $2$};

    \draw[semithick,<->,>=stealth] (4.8,2.1)-- +(0,1.4) node[midway, right] {\footnotesize $B\sub{c}$};
    \draw[semithick,<->,>=stealth] (0,2.05) -- +(4.75,0) node[midway, anchor=north] {\footnotesize $T\sub{c}$};

%
    
    \draw[semithick,<->,>=stealth] (2,6.6) -- +(1,0) node[midway, anchor=north] {\scriptsize $d$};
    \draw[semithick,<->,>=stealth] (1.85,6.7) -- +(0,0.6) node[midway, anchor=east] {\scriptsize $u$};    
    \node[anchor=south] at (2.5,7.3) {\scriptsize  RB};

    
\end{tikzpicture}
\caption{Two rounds of \gls{harqir} transmission with parameters $u=6$, $d=5$, $L\sub{c}=4$, $L=2$, and $\nmax=2$.}
 \label{fig:HARQ}
\end{figure}
We consider \gls{ofdm}  and assume that a \gls{rb} consists of~$d$ \gls{ofdm} symbols, each spanning $u$ subcarriers.
Therefore, an \gls{rb} consists of~$\nc = ud$ symbols.
The time interval over which an \gls{rb} is transmitted is referred to as slot, in accordance with the \gls{3gpp} terminology.
We let $L$ be the number of diversity branches used in a slot out of the available $L\sub{c}$.
Hence, $L$ \glspl{rb}, amounting to $L\nc$ symbols, are transmitted in each slot, as shown in Fig. \ref{fig:FBL}.

We shall consider two communication schemes: \gls{fbl} and \gls{harqir}.
In the \gls{fbl} case, a codeword is transmitted over $L$ diversity branches per slot and $v$ slots in total.
Hence, a codeword consists of~$v L \nc$ symbols, see Fig.~\ref{fig:FBL}.
We assume that transmissions on different slots occur on different diversity branches, so that transmitted symbols in different slots experience independent fading. 
This means that the number of slots per codeword is no larger than $\nmax =  \floor{L\sub{c} / L}$.
At the receiver side, decoding is performed once the entire codeword is received.

In the \gls{harqir} case, we consider one direction of a bidirectional communication link.
We assume that communication is organised in rounds, each consisting of two consecutive slots within the same diversity branches (time-division duplexing).
The two communicating devices are assigned one slot each in every round.
We assume that transmissions on different rounds occur on different diversity branches (see Fig. \ref{fig:HARQ}).
Hence, transmitted symbols in different rounds experience independent fading.
Consequently, the maximum number of rounds allowed is $\nmax$.
In each round, a device transmits $L\nc$ symbols during one slot and listens for an ACK/NACK and possibly other data during the next slot.
At the other device (receiver), decoding is attempted  after the first slot.
In the second slot, the receiver piggybacks an ACK/NACK on its data, to inform the transmitter whether to continue or to terminate transmission.
If a NACK is piggybacked, incremental redundant symbols are sent during the next round.
Rounds go on until the transmitter observes an ACK or it reaches the maximum predetermined number of rounds.
Such an ACK/NACK transmission introduces a delay compared to \gls{fbl} transmission, which we will take into account in our analysis.
We  assume that the feedback delay per round amounts to one slot.
However, our analysis can be easily generalized to arbitrary feedback delays.
As illustrated in Fig.~\ref{fig:HARQ}, when the same amount of resources are used in \gls{harqir} (per communication direction) and \gls{fbl}, the \gls{fbl} scheme requires half of the time of the \gls{harqir} scheme.

The input-output relation for the slot assigned for the forward transmission in round $j$, satisfies
\begin{IEEEeqnarray}{rCl}
\label{eq:sys_mod1}
P_{\randvecy_j \given \randvecy_1, \dots \randvecy_{j-1}, \randvecx_1, \dots, \randvecx_j} &=& P_{\randvecy_{j} \given \randvecx_{j}}=\prod_{k=1}^L P_{\randvecy_{j,k}\given \randvecx_{j,k}}
\end{IEEEeqnarray}
with
\begin{IEEEeqnarray}{rCl}
\label{eq:sys_mod2}
\randvecy_{j,k} = H_{j,k} \randvecx_{j,k} + \randvecw_{j,k}.
\end{IEEEeqnarray}
Here, ${\randvecy_{j} = {\lrho{\randvecy_{j,1}, \dots, \randvecy_{j,L}}}}$ and ${\randvecx_j = {\lrho{\randvecx_{j,1}, \dots, \randvecx_{j,L}}}}$, where ${\randvecy_{j,k} \in \complexset^{\nc}}$ and ${\randvecx_{j,k} \in \mathcal{X} = \lrbo{\vecx \in \complexset^{\nc} : \vecnorm{\vecx}^2 = \nc\rho}}$ for $j=1,2,\dots$, and $k=1,\dotsc, L$.
The variable $\rho$ denotes the SNR.
The Rayleigh fading is modeled by $H_{j,k} \distas \cgauss{0}{1}$, and $\randvecw_{j,k} \distas \cgauss{\bm{0}}{\rmatI_{\nc}}$ is the AWGN noise.
The random variables $\lrb{H_{j,k} }$ and $\lrb{\randvecw_{j,k} }$, which are mutually independent, are also independent over $j$ and $k$.
No \emph{a priori} knowledge of the realizations of $\lrbo{H_{j,k}}$ is available at the transmitter and at the receiver.
Consequently, it is reasonable to transmit equal-power signals over the available diversity branches.
This justifies our assumption that $\randvecx_{j,k} \in \mathcal{X}$ for all $j$ and $k$.

A code for the channel \eqref{eq:sys_mod1} in the \gls{harqir} setup is formally defined next by adapting the notion of \gls{vlsf} codes in \cite{polyanskiy2011feedback}.
\begin{dfn}\label{def:code}
	An $\lro{\ell, M, \epsilon, \rho, \nmax}$-\gls{vlsf} code, where $\ell \geq 1$, $M$, $\nmax$ are positive integers, $\rho >0$, and $0 < \epsilon < 1$, consists of
	\begin{itemize}
	\item[1)] A random variable $U$ with distribution $P_{U}$ defined on a set $\mathcal{U}$ with $\abs{\,\mathcal{U}}\leq 2$ that is revealed to both the transmitter and the receiver before the start of transmission. $U$ acts as a common randomness and enables the use of randomized encoding and decoding strategies.

	\item[2)] An encoder $f: \mathcal{U} \times \lrbo{1,\dotsc,M} \rightarrow \mathcal{X}^{L\nmax}$, that maps a message $J$, which is uniformly distributed on $\lrbo{1,\dotsc,M}$, to a codeword in the set $\lbrace\vecc(1),\dotsc,\vecc(M)\rbrace$.
	Each codeword is structured as $\vecc\lro{m} = \lrho{\vecc_1(m), \dots, \vecc_{\nmax}(m)}$ where $\vecc_j(m)\in\mathcal{X}^L$ for $j=1,\dotsc, \nmax$ and $m=1, \dotsc, M$.

	\item[3)] A sequence of decoders $g_v:\mathcal{U} \times \complexset^{\nc L v} \rightarrow \lrbo{1,\dotsc,M}$, $1\leq v \leq \nmax$, and a stopping time $\tau^* $, that is adapted to the filtration $\lrbo{\sigma\lro{U,\randvecy_1, \dots, \randvecy_v}}_{v=1}^{\nmax}$, and satisfies both
	\begin{IEEEeqnarray}{rCl}
		\Ex{}{\tau^*} \leq \ell,
	\end{IEEEeqnarray}
	and the average packet error probability target
	\begin{IEEEeqnarray}{rCl}
 \prob{ g_{\tau^*}\lro{U,\randvecy_1,\dotsc,\randvecy_{\tau^*}}\neq J } \leq \epsilon.
	\end{IEEEeqnarray}
	\end{itemize}
\end{dfn}

After the stopping time is triggered, the decoder uses the feedback channel to inform the encoder, through a one-bit ACK, to stop the transmission of the current message and to move to the transmission of the next one.
We  assume throughout the paper that this bit is error free and that so are the NACK bits transmitted in the previous feedback rounds.
Differently from most literature, we will consider in our analysis undetected error events in which an ACK is fed back although the decoder has chosen the wrong message.


For a given $\ell$, $\epsilon$, and $\nmax$, the maximum coding rate $R^*$, measured in information bits per channel use is defined as
\begin{IEEEeqnarray}{rCl}
R^*\lro{\ell, \epsilon,\rho, \nmax} &=& \nonumber \\
 && \hspace{-3cm}\sup\lrbo{\frac{\log_2 M}{\ell L\nc} : \exists \lro{\ell, M, \epsilon, \rho,  \nmax}\text{-\gls{vlsf} code}}.\label{eq:R*}
\end{IEEEeqnarray}

We will also be interested in the problem of minimizing the average number of transmissions $\ell$ for a given number of messages $M$, which yields the following definition:
\begin{IEEEeqnarray}{rCl} \label{eq:min_eps}
  \ell^*(M, \epsilon, \rho, \nmax) &=& \nonumber  \\
  &&\hspace{-2cm} \inf\lrbo{\ell : \exists \lro{\ell, M, \epsilon,\rho,  \nmax}\text{-\gls{vlsf} code}}.
\end{IEEEeqnarray}%
Some of our results will also be expressed in terms of the minimum energy per bit\footnote{We will not consider the energy spent to send the feedback bit reliably.} $\Emin$, which is related to $\ell^*$ as 
\begin{equation}\label{eq:minimum_energy_per_bit}
  \Eminf = \frac{\rho  L\nc}{\log_2 M}\,\ell^*(M, \epsilon, \rho, \nmax) .
\end{equation}
The corresponding metrics for \gls{fbl} are defined as in \cite{johan:unpublished}.

\subsection{PAT with SNN Decoding}
\label{sec:pat_and_snn_decoding}
Following \cite{johan:unpublished}, we assume that, for slot $j$ and coherence interval $k$, the input vector $\randvecx_{j,k}$ is of the form $[\xp, \xd_{j,k}]$, where $\vecx^{\lro{\text{p}}} \in \complexset^{\np}$, $1\leq \np < \nc$, is a deterministic vector containing pilot symbols and $\vecnorm{\vecx^{\lro{\text{p}}}}^2 =\np\rho$.
The vector $\xd_{j,k} \in \complexset^{\nd}$ contains the ${\nd=\nc-\np}$ data symbols.
Let $P_{\xd}$ denote a distribution on $\complexset^{\nd}$ such that $\randvecx_{j,k} \in \mathcal{X}$ \wpone when $\xd_{j,k} \sim P_{\xd}$.
Let $\yp_{j,k}$ and $\yd_{j,k}$ be the vectors containing the received samples corresponding to the pilot and data symbols, respectively. %
Given $\yp_{j,k}$ and $\yd_{j,k}$, the receiver computes the maximum likelihood estimate $\widehat{H}_{j,k}$ of the fading coefficient $H_{j,k}$ as
\begin{IEEEeqnarray}{rCl} \label{eq:decoding_rule}
	\widehat{H}_{j,k} = \frac{1}{\np \rho} \herm{(\vecx^{\lro{\text{p}}})} \yp_{j,k}.
\end{IEEEeqnarray}

Next we define the \gls{snn} decoder.
Denote a candidate codeword and the observed channel outputs up to the $v$th slot by~${\vecx^v=\lrho{\vecx_{1,1},\dotsc, \vecx_{1,L},\dotsc, \vecx_{v,L}}}$ and~${\vecy^v  =\lrho{\vecy_{1,1},\dotsc,\vecy_{1,L},\dotsc, \vecy_{v,L}}}$, respectively, where $\vecx_{j,k} \in \mathcal{X}$ and $\vecy_{j,k} \in \complexset^{\nc}$ for $j=1,\dotsc,v$, and $k=1,\dotsc, L$.
Here, $\vecx_{j,k}$ and $\vecy_{j,k} $ follow the structure outlined above.
The \gls{snn} metric is given by
\begin{equation}\label{eq:decoding_metric}
  q^{(v)}(\vecx^v, \vecy^v) = \prod_{j=1}^{v} \prod_{k=1}^L q(\vecx_{j,k},\vecy_{j,k})
\end{equation}
where
\begin{IEEEeqnarray}{rCl}
	q(\vecx_{j,k},\vecy_{j,k}) = \prod_{i=1}^{\nd} \exp\lro{-\abs{y^{\lro{\text{d}}}_{j,k,i} -  \widehat{h}_{j,k}x_{j,k,i}^{\lro{\text{d}}} }^2}.\IEEEeqnarraynumspace \label{eq:qmetric}
\end{IEEEeqnarray}
Here, $y^{\lro{\text{d}}}_{j,k,i}$ and $x^{\lro{\text{d}}}_{j,k,i}$ denote the $i$th entry of $\vecy^{\lro{\text{d}}}_{j,k}$ and $\vecx^{\lro{\text{d}}}_{j,k}$, respectively.
We  refer to this coding scheme as \gls{pat}-\gls{snn}.

Finally, we define the generalized information density in slot~$v$ as a mapping~$\mathcal{X}^{vL} \times \complexset^{\nc L v}  \rightarrow \reals$, given as
\begin{IEEEeqnarray}{rCl}\label{eq:inf_dens_acc}
\imath_{s}^v\lro{\vecx^v,\vecy^v} &=& \sum_{j=1}^v \sum_{k=1}^{L} \imath_{s}\lro{\vecx_{j,k},\vecy_{j,k}}
\end{IEEEeqnarray}
where the generalized information density per coherence block $\imath_{s}(\vecx_{j,k}, \vecy_{j,k})$ is
\begin{IEEEeqnarray}{rCl}\label{eq:inf_dens}
\imath_{s}(\vecx_{j,k},\vecy_{j,k}) &= &\log\frac{q(\vecx_{j,k},\vecy_{j,k})^{s}}{\Ex{\overline{\randvecx}}{q(\overline{\randvecx},\vecy_{j,k})^{s}}}.
\end{IEEEeqnarray}
Here, $\overline{\randvecx}=[{\vecx}^{\lro{\text{p}}},\matdatabar]$ with ${\vecx}^{\lro{\text{p}}}$ an arbitrary pilot vector satisfying the properties listed above, $\matdatabar \sim  P_{\xd}$, and $s\geq 0$.
For the special case in which $P_{\xd}$ is a product distribution, i.e., $P_{\xd}(\xdd_{j,k}) = \prod_{i=1}^{\nd} P_{X}(x^{\lro{\text{d}}}_{j,k,i})$, we can write~\eqref{eq:inf_dens} as
\begin{IEEEeqnarray}{rCl}\label{eq:prod_inf_dens}
\imath_{s}(\vecx_{j,k},\vecy_{j,k}) &=&
\sum_{i=1}^{\nd} -s\abs{y^{\lro{\text{d}}}_{j,k,i} -  \widehat{h}_{j,k}x^{\lro{\text{d}}}_{j,k,i}  }^2 \nonumber \\
 &\>& - \log\Ex{\overline{X}}{\exp\lro{-s\abs{y^{\lro{\text{d}}}_{j,k,i}-  \widehat{h}_{j,k} \overline{X} }^2}} .\IEEEeqnarraynumspace
\end{IEEEeqnarray}

\section{Finite-Blocklength Achievability Bounds} 

\subsection{Fixed-Blocklength Transmission without Feedback}
\label{sec:singe_shot_transmission}
Next, we review an achievability bound for the channel in~\eqref{eq:sys_mod2}, based on the random-coding union bound with parameter $s$~\cite[Thm. 1]{martinez11-02a}.
This bound will be used to assess the performance of \gls{pat}-\gls{snn}-\gls{fbl} transmission.

\begin{thm}\label{thm:RCUs_general}
Fix an integer $1\leq\nd<\nc$, a rate $R$, and a real number $s\geq 0$.
The average error probability for the \gls{pat}-\gls{snn}-\gls{fbl} scheme, operating as described in Section~\ref{sec:pat_and_snn_decoding}, is upper-bounded as
\begin{IEEEeqnarray}{rCl}\label{eq:RCUs_error}
 \epsilon &\leq &  \Ex{}{\exp\lro{-\lrho{\imath^v_{s}(\randvecx^v,\randvecy^v)-\log(2^{v L\nc R}-1) }^{+}} } \IEEEeqnarraynumspace
\end{IEEEeqnarray}
where ${\randvecx^v=\lrho{\randvecx_{1,1},\randvecx_{1,2},\dotsc, \randvecx_{L,1},\dotsc, \randvecx_{v,L}}}$ and each $\randvecx_{j,k}=[\xp, \xd_{j,k}]$ with $\xp$ an arbitrary pilot vector and~$\xd_{j,k} \distas P_{\xd}$, and $\randvecy_{j,k}$ is the channel output according to~\eqref{eq:sys_mod2}, for $j=1,\dotsc,v$ and $k=1,\dotsc,L$.
\end{thm}

\begin{IEEEproof}
See \cite[Thm. 3]{johan:unpublished}.
\end{IEEEproof}
\paragraph*{Remark} The latency for the \gls{fbl} scheme is  $T\sub{d} = v d T\sub{o}$, where $T\sub{o}$ is the \gls{ofdm} symbol duration.
In the \gls{harqir} case we will analyze next, latency is a random variable.
We will investigate both its average value and its \gls{cdf}.
\subsection{\gls{harqir} Transmission}
\label{sec:harq_transmission}
We provide next an achievability bound for the \gls{harqir} setup, which closely follows~\cite[Thm. 3]{polyanskiy2011feedback}.
To prove~\cite[Thm. 3]{polyanskiy2011feedback}, one computes the accumulated information density for each codeword and stops when one of the information densities exceeds a given threshold.
In our case, instead of information density, we accumulate generalized information density, see~\eqref{eq:inf_dens}.
Furthermore, we only allow for a fixed number of transmission rounds, after which transmission is terminated and a decision is taken.
This differs from the setup in \cite{williamson15-07a} where, after a fixed number of transmission rounds, the received data is discarded and transmission is restarted.
Incorporating these changes to~\cite[Thm. 3]{polyanskiy2011feedback}, one obtains the following result.

\begin{thm}[Achievability \gls{harqir}]\label{thm:HARQ_ach}
Fix three scalars~$\gamma>0$, $s \geq0$, $\rho > 0$, and two positive integers $\nmax$ and $1\leq \np < \nc$.
Let~$\lrbo{\randvecx_j}_{j=1}^{\infty}$ be a stochastic process such that~$\randvecx_j = \lrho{\randvecx_{j,1},\randvecx_{j,2},\dotsc,\randvecx_{j,L}}$, and
$\randvecx_{j,k}\in \mathcal{X}$ are \gls{iid}  for $j = 1,\dots$ and $k=1\dots,L$.
Let $\randvecx_{j,k} = [{\vecx}^{\lro{\text{p}}}, \xd_{j,k}]$ where ${\vecx}^{\lro{\text{p}}}$ is an arbitrary pilot vector satisfying $\vecnorm{\xp}^2=\np\rho$ and $\xd_{j,k}\distas P_{\xd}$.
Let~$\lbrace\overline{\randvecx}_j\rbrace_{j=1}^{\infty}$ be an independent copy of~$\lrbo{\randvecx_j}_{j=1}^{\infty}$.
Moreover, define the stopping times
\begin{IEEEeqnarray}{rCl}
{\tau} &=& \inf\{v\geq 1 :\imath_{s}^v\lro{\randvecx^v,\randvecy^v} \geq\gamma\},\\
\overline{{\tau}} &=& \inf\{v\geq 1 :\imath_{s}^v({\overline{\randvecx}^v,\randvecy^v}) \geq\gamma\}, \label{eq:taubar}
\end{IEEEeqnarray}
where $\randvecy^v$ is the random vector representing the channel outputs corresponding to the channel input $\randvecx^v$ between the first and the $v$th round, according to \eqref{eq:sys_mod1}.
Then, there exists an~$\lro{\ell, M, \epsilon, \rho, \nmax}$-\gls{vlsf} code with
\begin{IEEEeqnarray}{rCl}
\ell &\leq& \Ex{}{\min\lrbo{\nmax,{\tau}}}, \label{eq:ell_def}\\
\epsilon &\leq& \lro{M-1} \prob{\overline{\mathbf{\tau}}\leq\min\lrbo{\nmax,\mathbf{\tau}}}+\prob{\mathbf{\tau}> \nmax}. \IEEEeqnarraynumspace \label{eps_feedback_def}
\end{IEEEeqnarray}
\end{thm}
\begin{IEEEproof}
We create a codebook with $M$ independent codewords, each belonging to $\mathcal{X}^{L\nmax}$.
Each codeword is a composition of $L\nmax$ subcodewords, belonging to $\mathcal{X}$, where each subcodeword contains $\np$ deterministic pilot symbols and $\nd=\nc-\np$ data symbols distributed according to $P_{\xd}$.

For the $m$th codeword, $1\leq m \leq M$, we define a stopping time based on the generalized information density in \eqref{eq:inf_dens_acc} as
\begin{IEEEeqnarray}{rCl}
	\tau_m \!=\! \min \lrbo{\inf\lrbo{1\! \leq \! v \! \leq\!  \nmax : i_s^v\lro{\vecc^{v}(m), \randvecy^v} \geq \gamma}, \nmax} \IEEEeqnarraynumspace
\end{IEEEeqnarray}
where we use the convention that the infimum of the empty set equals infinity. 
Let now $\tau^* = \min\lrbo{\tau_1,\tau_2,\dotsc,\tau_M}$.
Note that this stopping time is different from the one in~\cite[Thm. 3]{polyanskiy2011feedback} since we are dealing here with a finite number of transmission.
Let
\begin{IEEEeqnarray}{rCl}
\hat{J}_{\tau^*}=\argmax_{1\leq m \leq M} \lrbo{ i_s^{\tau^*}(\vecc^{\tau^*} \!(m), \randvecy^{\tau^*} ) }
\end{IEEEeqnarray}
denote the estimated message at stopping time $\tau^*$.
In words, if the threshold is crossed within $\nmax$ transmission rounds, we choose the message corresponding to the largest overshoot.
If the threshold is not crossed within $\nmax$ transmission rounds, we simply choose the message with the largest accumulated metric.
For this decoding scheme, the average stopping time, averaged over the codebook ensemble generated by the distribution $P_{\xd}$, is upper-bounded as
\begin{IEEEeqnarray}{rCl}
\Ex{}{\tau^*} 
&\leq& \frac{1}{M}\sum_{j=1}^{M} \Ex{}{\tau_j\given J=j}
=  \Ex{}{\min\lrbo{{\tau}, \nmax}}\label{eq:ens_av_1}.
\end{IEEEeqnarray}
In words, we upper-bound $\tau^*$ by the minimum between $\nmax$ and the number of rounds required for the accumulated generalized information density of the transmitted codeword to exceed the threshold.
Next we define two error events
\begin{IEEEeqnarray}{rCl}
\mathcal{E}_1\! &=&\! \lbrace \tau^*\! \leq \!\nmax, \hat{J}_{\tau^*} \neq J, \max_m \lbrace i_s^{\tau^*}(\vecc^{\tau^*} \!(m), \randvecy^{\tau^*})\rbrace \!\geq\! \gamma \rbrace, \IEEEeqnarraynumspace\\
\mathcal{E}_2\! &=&\! \lbrace \tau^* \! =\! \nmax, \hat{J}_{\tau^*} \neq J, \max_m\lbrace i_s^{\tau^*}(\vecc^{\tau^*} \!(m), \randvecy^{\tau^*})\rbrace \!<\! \gamma \rbrace,\IEEEeqnarraynumspace
\end{IEEEeqnarray}
such that $\epsilon = \prob{ \mathcal{E}_1 \cup \mathcal{E}_2 }$.
Here, $\mathcal{E}_1$ is the event the threshold is crossed within $\nmax$ transmission rounds and an erroneous message is chosen, and $\mathcal{E}_2$ is the event that, in transmission round $\nmax$, the threshold has not been crossed and the decoder decides for the wrong message.
Hence, $\epsilon$ describes the undetected error event.
We proceed as follows:
\begin{IEEEeqnarray}{rCl}
\epsilon &\leq & \prob{\mathcal{E}_1} + \prob{\mathcal{E}_2} \\
&\leq &
\sum_{j=2}^{M}\prob{\tau_j \!\leq\! \tau_1, i_s^{\tau_j}\lro{\vecc^{\tau_j}\lro{j}, \randvecy^{\tau_j}} \!\geq\! \gamma \given J=1} + \prob{{\tau}\!>\! \nmax} \nonumber \\
&=&    \lro{M-1}\prob{\overline{{\tau}}\leq\min\lrbo{\nmax,\tau}} + \prob{{\tau} \!>\! \nmax} \label{eq:ens_av_2}.
\end{IEEEeqnarray}
%
%
%
Using the arguments in~\cite[Thm. 19]{polyanskiy2011feedback} and~\cite[p. 35]{Eggleston58}, we conclude that a randomized codebook attaining both ensemble averages \eqref{eq:ens_av_1} and \eqref{eq:ens_av_2} simultaneously can be constructed by using a convex combination of two deterministic codebooks.
\end{IEEEproof}
\paragraph*{Remark} The average latency for the \gls{harqir} scheme is $T\sub{d} = 2\ell d T\sub{o}$, where the factor $2$ is due to the feedback delay and $T\sub{o}$ is the \gls{ofdm} symbol duration.
The maximum latency is, however, $ 2 \nmax d T\sub{o}$.
Note also that the \gls{cdf} of the delay $\tau$ is required to compute~\eqref{eq:ens_av_2}.

No closed-form expressions for \eqref{eq:ell_def} and \eqref{eps_feedback_def} are available.
Furthermore, evaluating \eqref{eps_feedback_def} is numerically challenging.
Indeed, $M$ is typically very large (for a fixed rate $R$, it grows exponentially in the blocklength), so the first probability is very small thus preventing the use of Monte Carlo methods.
For the case of summands in \eqref{eq:taubar} with negative drift, i.e., $\Ex{}{\imath_{s}(\overline{\randvecx}_{j,k},\randvecy_{j,k})} < 0$, we can further relax \eqref{eps_feedback_def} by upper-bounding $\prob{\overline{\mathbf{\tau}}\leq\min\lrbo{\nmax,\mathbf{\tau}}} $ using Wald's identity as \cite[Cor. 9.4.4]{gallager13}
\begin{IEEEeqnarray}{rCl}
	 \prob{\overline{\mathbf{\tau}}\leq\min\lrbo{\nmax,\mathbf{\tau}}}&\leq &  \prob{\overline{\mathbf{\tau}}\leq \infty}
	\leq \exp\lro{-\beta^* \gamma} \label{eq:relax}
\end{IEEEeqnarray}
where $\beta^*$ is the positive solution of
\begin{IEEEeqnarray}{rCl}
\kappa\lro{\beta} = \log \Ex{}{\exp\lro{\beta \sum_{k=1}^L \imath_{s}(\overline{\randvecx}_{1,k},\randvecy_{1,k})}} = 0.
\end{IEEEeqnarray}

In the upcoming section, we let $P_{\randvecx^{\lro{\text{d}}}}$ be a product distribution. 
Hence, the generalized information density is given by \eqref{eq:prod_inf_dens}.
It then follows from Jensen's inequality that $\Ex{}{\imath_{s}\lro{\overline{\randvecx}_{j,k},\randvecy_{j,k}}} < 0$. 
This enables us to use \eqref{eq:relax} in our numerical evaluations.
Furthermore, we can express $\kappa\lro{\beta}$ as
\begin{IEEEeqnarray}{rCl}
	\kappa\lro{\beta} &=&
	L \log \Ex{}{ \frac{\Ex{}{q\lro{\overline{\randvecx},\vecy}^{\beta s} \given \randvecy=\vecy}}{\Ex{}{ q\lro{\overline{\randvecx}, \vecy}^s\given \randvecy=\vecy}^{\beta}}	 }\IEEEeqnarraynumspace \label{eq:cgf}
\end{IEEEeqnarray}
where $\overline{\randvecx} = \lbrack \vecx^{\lro{\text{p}}}, \overline{\randvecx}^{\lro{\text{d}}}\rbrack$ with $\overline{\randvecx}^{\lro{\text{d}}} \sim P_{\randvecx^{\lro{\text{d}}}}$ and $\randvecy$ distributed as in \eqref{eq:sys_mod2}, independent of $\overline{\randvecx}$.
From \eqref{eq:cgf}, we obtain $\beta^*=1$.
\paragraph*{Remark} An upper bound similar to~\eqref{eq:relax} with $\beta^*=1$ is obtained in~\cite[Eq. (113)]{polyanskiy2011feedback} by using a change-of-measure argument.
We cannot follow the same strategy here because of the mismatched decoding metric.

\section{Numerical Results}
\label{sec:numerical_results}
In this section, we compare the performance of \gls{harqir} and \gls{fbl}.
For the sake of concreteness, we consider \gls{iid} input symbols drawn uniformly from a \gls{qpsk} constellation.
We extract our channel parameters from the TDL-C 300 ns--3km/h channel model~\cite{3GPP-TR-38.901}.
We choose the system bandwidth, subcarrier bandwidth, and \gls{ofdm} symbol duration equal to $20$~MHz, $15$~kHz, and $71.4$~$\mu$s, respectively.
This results in $30$ available diversity branches.
The channel model has coherence bandwidth $B\sub{c}=0.66$~MHz, which implies $u \leq 44$.
We let $u=24$ and $d=3$. %
Hence, an \gls{rb} occupies $360$~kHz and its duration is $214.2$~$\mu$s.
We also let $\epsilon=10^{-3}$, which corresponds to the least stringent reliability constraint for \gls{urllc}~\cite{Schulz17}.
The latency for \gls{harqir} is $T\sub{d} = 2T\sub{o} d \ell$, and thus the minimum and maximum latency are $2T\sub{o} d$ and $2 T\sub{o} d \nmax$, respectively.
For \gls{fbl}, the minimum latency corresponds to one slot duration, i.e., $T\sub{o} d$, and the maximum latency is $T\sub{o} d \nmax$.
Throughout this section, we assume that each message contains $k=\log_2 M=30$ information bits.
The channel and system parameters are summarized in Table \ref{tab:channel-parameters}.

\begin{table}[tb]\centering
\caption{Parameters for TDL-C 300\,ns--3\,km/h (upper half) and signal (bottom half).}
\label{tab:channel-parameters}
\begin{tabular}{C{0.95cm}C{4.5cm}C{1.2cm}}
\toprule
\bf Symbol 			& \bf Parameter		&   Value\\
\midrule
$B\sub{c}$			& $50$\% coherence bandwidth 			& $0.66$ MHz \\
$T\sub{c}$			& $50$\% coherence time 				& $85$ ms \\
$L\sub{c}$		& Number of available diversity branches 			& $30$ \\
\midrule
$k$ & Information bits & 30\\
$\epsilon$ & Error probability & $10^{-3}$ \\
$B$ & System bandwidth & $20$ MHz \\
$B\sub{s}$ & Subcarrier bandwidth & $15$ KHz \\
$T\sub{o}$ & \gls{ofdm} symbol duration & 71.4 $\mu$s \\
$u$			& Subcarriers per \gls{rb}			& $24$ \\
$d$			& \gls{ofdm} symbols per \gls{rb}		& $3$ \\
$L$ & Number of used diversity branches & $\leq L\sub{c}$ \\
$\nmax$ & Max. number of transmission rounds & $\floor{L\sub{c} / L}$\\
\bottomrule
\end{tabular}
\end{table}

Next, we detail how we evaluate the minimum energy per bit~\eqref{eq:minimum_energy_per_bit}. The procedure for computing the maximum coding rate~\eqref{eq:R*} is similar.
In both \gls{fbl} and \gls{harqir}, we fix $\rho$, $\epsilon$, $k$, $T\sub{o}$, $u$, $d$, and $L$.
For \gls{harqir}, we find the smallest threshold $\gamma$ such that the right-hand side of~\eqref{eps_feedback_def}, relaxed using \eqref{eq:relax}, is below the desired error target.
This $\gamma$ is then used to evaluate  \eqref{eq:ell_def}, which provides an upper bound on the average latency $T\sub{d}$.
For \gls{fbl}, we search for the smallest $v$ such that the right-hand side of~\eqref{eq:RCUs_error} is below the desired error target.
Throughout this section, all charts are obtained by optimizing over the parameter~$s$ and the number of pilot symbols $\np$.

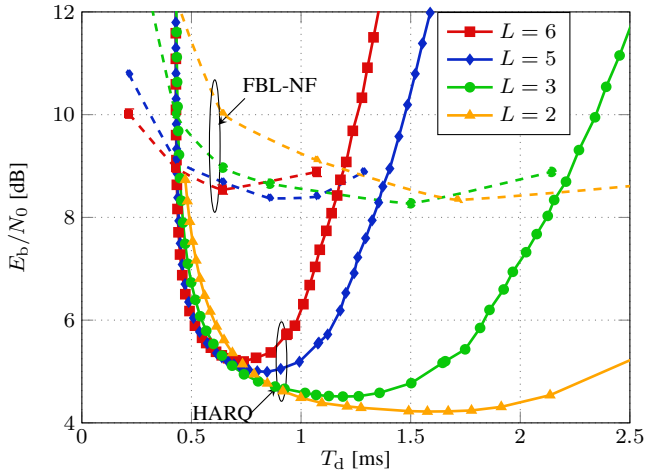
\begin{figure}[t]
\centering
\begin{tikzpicture}[font=\footnotesize, scale=1]
\begin{axis}[
       xmin = 0,
       xmax = 3,
       xlabel = $\ell$,
       xticklabels={ {$1$}, {$2$}, {$3$},{$4$},{$5$},{$6$},{$7$},{$8$},{$9$},{$10$}},
       xtick={0.4284 ,   0.8568 ,   1.2852  ,  1.7136  ,  2.1420 ,  2.5704  ,  2.9988 ,   3.4272   , 3.8556 ,   4.2840},
       axis x line*=top,
       xlabel near ticks,
       grid=both,
        hide y axis,
         hide x axis,
              legend style={at={($(0.88,0.148)$)},legend columns=1, fill=white,draw=black,anchor=center,align=center},
       ]
       \addplot [draw=none] table [y index={1}, x index = {0}, col sep=comma] {./Data/NF_Rate_vs_ell_L_2_SNR_2.csv};

 \end{axis}
 \begin{axis}
     [
      xlabel={$T\sub{d}$ [ms]},
      ylabel = {$E\sub{b}/N_0$ [dB]},
      grid=both,
      ymin = 4,
      ymax =12,
      xmin = 0,
      xmax = 2.5,
      legend style={at={($(0.77,0.85)$)},legend columns=1, fill=white,draw=black,anchor=center,align=center},
      cycle list name=colorfour,
     ]
\addplot+ [mark=square*,solid, mark size=1.5,line width=\linew] table [y index={1}, x index = {0}, col sep=comma] {./Data/F_EbN0_vs_ell_L_6.csv}coordinate[pos=0.25](pt1);\addlegendentry{$ L=6$}
\addplot+ [ mark=diamond*,solid, mark size=1.5,line width=\linew] table [y index={1}, x index = {0}, col sep=comma] {./Data/F_EbN0_vs_ell_L_5.csv};\addlegendentry{$ L=5$}
\addplot+ [ mark=*,solid, mark size=1.5,line width=\linew] table [y index={1}, x index = {0}, col sep=comma] {./Data/F_EbN0_vs_ell_L_3.csv};\addlegendentry{$ L=3$}
\addplot+ [mark=triangle*,solid, mark size=1.5,line width=\linew] table [y index={1}, x index = {0}, col sep=comma] {./Data/F_EbN0_vs_ell_L_2.csv}coordinate[pos=0.27](ut1);\addlegendentry{$L=2$}


     \coordinate (pt2) at ($(pt1) + (0pt, -8pt)$);
     \draw[rotate=0](pt2) ellipse  (2pt and 15pt);
     \coordinate (pt3) at ($(pt2)+ (-2pt,-10pt)$);
     \coordinate (pt4) at ($(pt3)+ (-10pt,-10pt)$);
     \draw[<-] (pt3)--(pt4) node at ($(pt4) + (-10pt,0pt)$) {\gls{harqir}};

\addplot+ [mark=square*,dashed, mark size=1.5,line width=\linew,smooth,tension=0.2] table [y index={1}, x index = {0}, col sep=comma] {./Data/NF_EbN0_vs_ell_L_6.csv}coordinate[pos=0.7](ut1);
\addplot+ [mark=diamond*,dashed, mark size=1.5,line width=\linew,smooth,tension=0.2] table [y index={1}, x index = {0}, col sep=comma] {./Data/NF_EbN0_vs_ell_L_5.csv};
\addplot+ [ mark=*,dashed, mark size=1.5,line width=\linew,smooth,tension=0.2] table [y index={1}, x index = {0}, col sep=comma] {./Data/NF_EbN0_vs_ell_L_3.csv};
\addplot+ [ mark=triangle*,dashed, mark size=1.5,line width=\linew,smooth,tension=0.2] table [y index={1}, x index = {0}, col sep=comma] {./Data/NF_EbN0_vs_ell_L_2.csv};

     \coordinate (pt2) at ($ (ut1) +  (0pt,+15pt)$);
     \draw[rotate=0](pt2) ellipse  (2pt and 25pt);
     \coordinate (pt3) at ($(pt2)+ (1.3pt,8pt)$);
     \coordinate (pt4) at ($(pt3)+ (+10pt,10pt)$);
     \draw[<-] (pt3)--(pt4) node at ($(pt4) + (13pt,5pt)$) {\gls{fbl}};

\end{axis}

 \end{tikzpicture}
 
 
\caption{$\Emin$ versus average latency for $L \in \lrbo{2,3,5,6}$ and $30$ information bits. The remaining parameter values are given in Table \ref{tab:channel-parameters}.}
\label{fig:sim_Eb}
\end{figure}

In Fig. \ref{fig:sim_Eb}, we show the minimum energy per bit required to fulfill the reliability target as a function of average latency.
It can be seen that, for fixed $L$, \gls{harqir} significantly outperforms \gls{fbl}.
For example, for $L= 3$ and $T\sub{d}=1$~ms, the difference is about $4$~dB.
It can also be seen that $\Emin$ does not decrease monotonically with the average latency for both \gls{harqir} and \gls{fbl}, but there exists a latency $T\sub{d}^*$ for which $\Emin$ is minimized.
When $T\sub{d}<T\sub{d}^*$, increasing the average blocklength provides an SNR gain that outweighs the  rate penalty (see \eqref{eq:minimum_energy_per_bit}), whereas the opposite is true when $T\sub{d}>T\sub{d}^*$.
\gls{harqir} performs better than \gls{fbl} because for a fixed $\epsilon$, changes in the average blocklength result in larger changes in the required SNR  than in \gls{fbl}.
We also observe a tradeoff between the average latency and the number of  diversity branches used: choosing a large $L$ is more energy efficient for small average latency ($T\sub{d} < 0.7$ ms in Fig.~\ref{fig:sim_Eb}), since in this regime the number of possible retransmissions is small, and thus higher diversity is beneficial. As the average latency increases, choosing a smaller number of diversity branches is preferable.

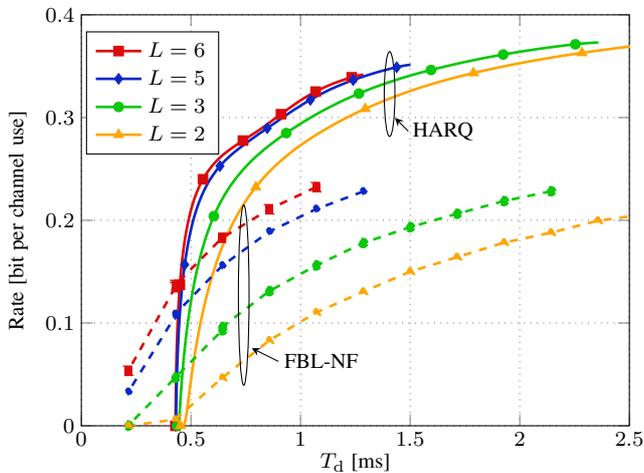
\begin{figure}[t]
\centering
\begin{tikzpicture}[font=\footnotesize]
\begin{axis}[
       xmin = 0,
       xmax = 3,
       xlabel = $\ell$,
       xticklabels={ {$1$}, {$2$}, {$3$},{$4$},{$5$},{$6$},{$7$},{$8$},{$9$},{$10$}},
       xtick={0.4284 ,   0.8568 ,   1.2852  ,  1.7136  ,  2.1420 ,  2.5704  ,  2.9988 ,   3.4272   , 3.8556 ,   4.2840},
       axis x line*=top,
       xlabel near ticks,
       grid=both,
        hide y axis,
         hide x axis,
       ]
       \addplot [draw=none] table [y index={1}, x index = {0}, col sep=comma] {./Data/NF_Rate_vs_ell_L_2_SNR_2.csv};

 \end{axis}
 \begin{axis}
     [
      xlabel={$T\sub{d}$ [ms]},
      ylabel = {Rate [bit per channel use]},
      grid=both,
      ymin = 0,
      ymax =0.4,
      xmin = 0,
      xmax = 2.5,
      legend style={at={($(0.13,0.81)$)},legend columns=1, fill=white,draw=black,anchor=center,align=center},
      cycle list name=colorfour 
      ]
\addplot+ [mark=square*,solid, mark size=1.5,line width=\linew,mark repeat={50}] table [y index = {1}, x index = {0}, col sep=comma] {./Data/F_Rate_vs_ell_L_6_SNR_-2.csv}coordinate[pos=0.6](ut1);\addlegendentry{$ L=6$}
\addplot+ [mark=diamond*,solid, mark size=1.5,line width=\linew,mark repeat={50}] table [y index = {1}, x index = {0}, col sep=comma] {./Data/F_Rate_vs_ell_L_5_SNR_-2.csv};\addlegendentry{$ L=5$}
\addplot+ [mark=*,solid, mark size=1.5,line width=\linew,mark repeat={50}] table [y index = {1}, x index = {0}, col sep=comma] {./Data/F_Rate_vs_ell_L_3_SNR_-2.csv};\addlegendentry{$ L=3$}
\addplot+ [mark=triangle*,solid, mark size=1.5,line width=\linew,mark repeat={50}] table [y index = {1}, x index = {0}, col sep=comma] {./Data/F_Rate_vs_ell_L_2_SNR_-2.csv}coordinate[pos=0.6](pt1);\addlegendentry{$ L=2$}

     \coordinate (pt2) at ($(pt1) !.6! (ut1)$);
     \draw[rotate=0](pt2) ellipse  (2pt and 16pt);
     \coordinate (pt3) at ($(pt2)+ (2pt,-9pt)$);
     \coordinate (pt4) at ($(pt3)+ (+5pt,-5pt)$);
     \draw[<-] (pt3)--(pt4) node at ($(pt4) + (13pt,0pt)$) {\gls{harqir}};
\addplot+ [mark=square*,dashed,  mark size=1.5,line width=\linew] table [y index={1}, x index = {0}, col sep=comma] {./Data/NF_Rate_vs_ell_L_6_SNR_-2.csv};

\addplot+ [mark=diamond*,dashed, mark size=1.5,line width=\linew] table [y index={1}, x index = {0}, col sep=comma] {./Data/NF_Rate_vs_ell_L_5_SNR_-2.csv}coordinate[pos=0.5](ut1);
\addplot+ [mark=*,dashed, mark size=1.5,line width=\linew] table [y index={1}, x index = {0}, col sep=comma] {./Data/NF_Rate_vs_ell_L_3_SNR_-2.csv};
\addplot+ [mark=triangle*,dashed,  mark size=1.5,line width=\linew] table [y index={1}, x index = {0}, col sep=comma] {./Data/NF_Rate_vs_ell_L_2_SNR_-2.csv}coordinate[pos=0.35](pt1);


     \coordinate (pt2) at ($(ut1) +  (0pt,-17pt)$);
     \draw[rotate=0](pt2) ellipse  (2pt and 34pt);
     \coordinate (pt3) at ($(pt2)+ (3pt,-20pt)$);
     \coordinate (pt4) at ($(pt3)+ (+10pt,-5pt)$);
     \draw[<-] (pt3)--(pt4) node at ($(pt4) + (16pt,0pt)$) {\gls{fbl}};
     
\end{axis}

 \end{tikzpicture}
 
 
\caption{$R^*$ versus average latency for $\rho=-2$ dB and $L\in\lrbo{2,3,5,6}$. The remaining parameter values are given in Table \ref{tab:channel-parameters}.}
\label{fig:sim_rate}
\end{figure}

In Fig. \ref{fig:sim_rate}, we compare the maximum coding rate achievable with the two schemes as a function of the average latency.
It can be seen that larger rates are achievable by using \gls{harqir} rather than \gls{fbl}.
A similar behavior was reported for the AWGN channel in~\cite{williamson15-07a}.
The rate increases with $L$ in both systems, since a larger $L$ allows for more diversity and longer codewords.

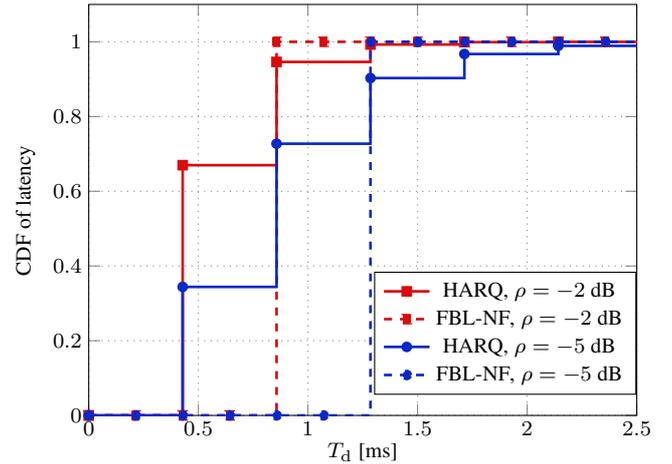
\begin{figure}[t]
\centering
\begin{tikzpicture}[font=\footnotesize, scale=1]
\begin{axis}[
       xmin = 0,
       xmax = 2.5,
       xlabel = $t$,
       xticklabels={ {$1$}, {$2$}, {$3$},{$4$},{$5$},{$6$},{$7$},{$8$},{$9$},{$10$}},
       xtick={0.4284 ,   0.8568 ,   1.2852  ,  1.7136  ,  2.1420 ,  2.5704  ,  2.9988 ,   3.4272   , 3.8556 ,   4.2840},
       axis x line*=top,
       xlabel near ticks,
       grid=both,
        hide y axis,
         hide x axis,
              legend style={at={($(0.8,0.5)$)},legend columns=1, fill=white,draw=black,anchor=center,align=center},
       ]
       \addplot [draw=none] table [y index={1}, x index = {0}, col sep=comma] {./Data/NF_Rate_vs_ell_L_2_SNR_2.csv};

 \end{axis}
 \begin{axis}
     [
      xlabel={$T\sub{d}$ [ms]},
      ylabel = {CDF of latency},
      grid=both,
      ymin = 0,
      ymax =1.1,
      xmin = 0,
      xmax = 2.5,
      legend style={at={($(0.76,0.2)$)},legend columns=1, fill=white,draw=black,anchor=center,align=center},
      cycle list name=colorfour,
     ]
\addplot [const plot,mark=square*,solid, mark size=1.5,line width=\linew, color=red] table [y index={1}, x index = {0}, col sep=comma] {./Data/F_CDF_L2_SNR_-2.csv}coordinate[pos=0.23](pt1);\addlegendentry{\gls{harqir}, $\rho=-2$ dB}


\addplot [ const plot,mark=square*,solid, mark size=1.5,line width=\linew, color=red,dashed] table [y index={1}, x index = {0}, col sep=comma] {./Data/NF_CDF_L2_SNR_-2.csv};\addlegendentry{\gls{fbl}, $\rho=-2$ dB}

\addplot [const plot,mark=*,solid, mark size=1.5,line width=\linew, color=blue] table [y index={1}, x index = {0}, col sep=comma] {./Data/F_CDF_L2_SNR_-5.csv}coordinate[pos=0.23](pt1);\addlegendentry{\gls{harqir}, $\rho=-5$ dB}


\addplot [ const plot,mark=*,solid, mark size=1.5,line width=\linew, color=blue, dashed] table [y index={1}, x index = {0}, col sep=comma] {./Data/NF_CDF_L2_SNR_-5.csv};\addlegendentry{\gls{fbl}, $\rho=-5$ dB}

\end{axis}

 \end{tikzpicture}
 
 
\caption{\gls{cdf} of the latency.
The curves are generated for $30$ information bits, $\rho=\lrbo{-5,-2}$ dB and $L=2$.
The remaining parameter values are given in Table \ref{tab:channel-parameters}.}
\label{fig:cdf}
\end{figure}

The results presented so far are in terms of average latency.
However, average and maximum latencies are not equal in \gls{harqir}.
Hence, one may argue that the above comparisons are unfair since the maximum latency with \gls{harqir} may be larger than the latency with \gls{fbl}.
In Fig. \ref{fig:cdf}, we illustrate the \gls{cdf} of the latency for both \gls{fbl} and \gls{harqir}, for the case $L=2$, and $\rho\in \lrbo{-5,-2}$ dB.
The \gls{cdf} of the \gls{fbl} latency is a step function since the latency is deterministic whereas the \gls{cdf} of \gls{harqir} is a staircase function with steps at multiples of the duration of a transmission round.
We see from Fig. \ref{fig:cdf}   that the probability of the latency in \gls{harqir} being larger than in \gls{fbl} decreases with SNR. 
For $\rho=-5$ dB, this probability is about $0.097$; for $\rho=-2$ dB, the probability is about $0.054$.
The reason is that, as the SNR increases, it is more likely that one round is enough to satisfy the reliability constraint.

An important application where the \gls{cdf} of the delay plays a pivotal role is in joint coding-queuing analyses of \gls{urllc} systems. 
Indeed, when taking into account also queuing delays, an optimum latency-aware design must be based on the \gls{cdf} of the latency rather than on  its average~\cite{rahul:unpublished}. 
As a consequence, the results in this paper are relevant to the joint coding-queuing analysis of \gls{urllc} systems over memoryless block-fading channels with limited diversity.
Specifically, the tools introduced in this paper can be used to extend the analysis in \cite{rahul:unpublished}, where the physical layer was modeled as a simple binary-input AWGN channel, to block-fading scenarios and practically-relevant pilot-based transmission schemes.

\bibliographystyle{IEEEtran}
\bibliography{./giubib}
\end{document}